\documentclass[10pt]{article}
\usepackage{color}
\usepackage{array,amsmath,booktabs}
\usepackage{amsmath}
\usepackage{amsthm}
\usepackage{amsfonts}
\usepackage{enumerate}
\usepackage{amssymb}
\usepackage{hyperref}
\usepackage{tabularx}
\usepackage{a4wide}
\hypersetup{colorlinks,linkcolor={blue},citecolor={blue},urlcolor={red}}
\title{How to choose a Compatible Committee?}
\author{Ritu Dutta\footnote{Department of Mathematics, Dibrugarh University, Dibrugarh; Email:ritudutta150@gmail.com}~~$\cdot$~Rajnish Kumar \footnote{Queen's Management School, Queen's University Belfast, UK; Email: rajnish.kumar@qub.ac.uk}~~$\cdot$~Surajit Borkotokey \footnote{Corr. Author: Department of Mathematics, Dibrugarh University, Dibrugarh; Email: sborkotokey@dibru.ac.in}} 

\newtheorem{theorem}{Theorem}

\newtheorem{exm}{Example}
\newtheorem{remark}{Remark}
\newtheorem{axiom}{Axiom}
\newtheorem{definition}{Definition}
\theoremstyle{plain}

\begin{document}
\maketitle
\begin{abstract}
\normalfont Electing a committee of size $k$ from $m$ alternatives ($k < m$) is an interesting problem under the multi-winner voting rules. However, very few committee selection rules found in the literature consider the coalitional possibilities among the alternatives that the voters believe that certain coalitions are more effective and can more efficiently deliver desired outputs. To include such possibilities, in this present study, we consider a committee selection problem (or multi-winner voting problem) where voters are able to express their opinion regarding interdependencies among alternatives. Using a dichotomous preference scale termed generalized approval evaluation we construct an $m$-person coalitional game which is more commonly called a cooperative game with transferable utilities. To identify each alternative's score we use the Shapley value (Shapley, 1953) of the cooperative game we construct for the purpose. Our approach to the committee selection problem emphasizes on an important issue called the compatibility principle. Further, we show that the properties of the Shapley value are well suited in the committee selection context too. We explore several properties of the proposed committee selection rule. 
\end{abstract}


\section{Introduction}\label{sec:1}
In this paper, we propose a committee selection rule based on the framework of cooperative game theory. Our rule takes care of the voters' perspectives regarding the compatibility of the committee members. Based on their beliefs on whether a committee sustains in the long run, we construct a cooperative game with transferable utilities or simply a TU game and obtain the Shapley value for each of the alternatives. If the committee to be selected is of size $k$, then the top $k$ ranked alternatives prescribed by the Shapley value will constitute the desired committee.\\
\normalfont In the committee selection or multi-winner election, we are given a set of alternatives, a set of voters, the preferences that each voter has over these alternatives, and a desired size $k$ of the committee to be elected (Lackner et al., $2022$). The goal is to select a committee of exactly $k$ alternatives based on the voters’ preferences. Choosing a subset of alternatives is much more complex than choosing a single best alternative (Ratliff et al., $2003$). Examples of such situations include selecting a set of movies for an airplane (Skowron et al., $2016$), choosing locations for common facilities (Shmoys et al., $1997$), short-listing alternatives for hiring a new team, or electing a parliament, etc (Elkind et al., $2017$). These examples differ in their nature and for each of them, we may expect that the selected group of alternatives would exhibit certain properties.  The simplest way to get a committee rule or multi-winner voting rule is to extend the single-winner voting rule by choosing the top $k$ alternatives. For this, we can adopt any of the $k$-plurality (Faliszewski et al., $2018$), $k$-Borda (Diss et al., $2016$; Kamwa et al., $2015$), $k$-approval (Aziz et al., $2016$; Kilgour et al., $2006$ ), etc. These rules used in the multi-winner elections are commonly called committee selection rules of size $k$. Following (Elkind et al., 2017) the class of committee selection rules where the members of the committee are chosen based on their excellence are known as the excellence-based committee rules. The other two main approaches to committee selection rules are based on diversity and proportionality. In the diversity-based committee, it is important to include at least one alternative in the committee from all possible groups of members. Similarly, proportionality-based committee rules are the ones in which members are selected proportionate to their share of votes. For a detailed review on committee selection rules, we refer to (Diss et al., $2016$; Barbera et. al., $2008$; Elkind et. al., $2017$; Skowron et al., 2016), etc.

None of these approaches, however, addresses the compatibility issue among the members of the elected committee that might influence their selection. Committee selection problems where compatibility is considered can be found in (Ratliff, $2006$; Uckelman, $2010$; Ratliff et al., $2014$; Darmann, $2013$), etc. Thus, there are committees that are purely based on individual excellence and others are purely based on group excellence. However, to survive in the long run, an ideal committee needs both individual and group excellence. Committees that are only based on individual excellence may not work smoothly or effectively in joint events. Similarly, committees that have only group excellence but whose members do not have individual competency may also suffer from serious problems. This may result, for example, in a complete dissolution of the committee due to too much individualism or pre-matured exit of a fraction of the committee members in between (Lucchetti et al., $2015$; Bernardi et al., $2019$). So, in order to achieve a more stable and an efficient committee, these two attributes, namely, individual excellence and group excellence should be combined in the committee selection process.   \\
\normalfont In our current paper, we mainly discuss a committee rule which gives more importance to individual excellence and compatibility among members as observed or believed by the voters rather than diversity, or proportionality. Our proposed committee selection rule includes the possibility of having inter-dependencies or binding behaviour among the members. The following three examples highlight our proposals in detail. 
\begin{exm}\label{ex:1}\rm
\normalfont Let there be three students $a$, $b$, and $c$ having skills in various project related works. Let student $a$ be better in literature review and analysis of the model outputs than student $b$. Let $c$ be competent with programming skills. For the successful completion of any project, all these skills are essential. However, in a specific project undertaken by a group of investigators, there is scope for recruiting only two students among them as research associates. Among the three students, the group will choose student $a$ and student $c$ (i.e., $(a, c)$), as they are better among all the possible combinations of alternatives namely, $(a, b)$ and $(b, c)$ in terms of their skills. However, suppose $a$ and $c$ may not be compatible with each other in groups as evidenced by their previous activities, and their joint effort is likely to harm the project. This is an example of a committee selection problem (or a problem of multi-winner voting) where voters should also look at the group compatibility of the committee members in addition to giving priority to their skills.
\end{exm}
\begin{exm}\label{ex:2}\rm\normalfont{(Moretti et al., $2012$)}
\normalfont Next, consider the problem of what to take on a backpacking trip in the mountains: a bottle of water (w), can be more essential (thus ranked higher) than a bottle of orange juice (o) or than a sandwich (s), where a bottle of orange juice could be preferred (e.g., for dietary reasons) to a sandwich. But if the problem is now which pair of the three is to put in the backpack, a bottle of water and a bottle of orange juice may be less preferred (because of the backpack weight) than a bottle of water and a sandwich together. The problem now is how the voters will evaluate the compatibility issues among the alternatives.
\end{exm}
\begin{exm}\label{ex:3}\rm
 \normalfont The third example is partially taken from (Lackner, et. al., $2022$) and (Uckelman, $2010$). We state the example in two parts. The first part is based on (Lackner et al., $2022$). Suppose that we are electing a three-seat committee from the five alternatives Alice (a), Bob (b), Charlie (c),
Dave (d), and Elaine (e). Let there be $100$ voters. And their preferences are given as follows. $66$ voters approve alternatives $a$, $b$, $c$, $33$ voters approve alternative $d$, and one voter approves alternative $e$. How to select a committee of size three? In particular, how to choose a compatible committee of size three $?$

\noindent The second part of the example is taken from (Uckelman, $2010$). Suppose further, that a sizeable number of voters believe the following two constraints:
\end{exm}
\begin{enumerate}
    \item Alice and Bob are the best alternatives, so any committee with one of them is better than any committee with neither, and
    \item Alice and Bob will fight if they are on the committee together, so any committee with both is worse than any committee with neither.
\end{enumerate}

The aforementioned three examples suggest that in order to ensure group compatibility of the members and also preserve some level of individual excellency in a committee, i.e., the assessment of their individual and joint capabilities (individual and coalitional worths) by the voters is necessary. 

\normalfont Cooperative game theory offers very useful insights to evaluate the individual and joint capabilities of the players in a group activity. The main motive for studying social situations under the cooperative game theoretic framework is to achieve a common goal under a binding agreement that otherwise requires more effort to achieve individually. Our hypothesis states that forming a committee is a common goal for alternatives based on voters' evaluation and the voters have common priors on the compatibility among the alternatives. Thus, our proposed voting situation can be modelled as a TU game where the Shapley value (Shapley, $1953$), one of the most popular solutions concepts of cooperative game theory provides the best possible committee after considering all alternative interactions. Thus, our proposed voting rule is Shapley value-based which is characterized by axioms of efficiency, null player property, symmetry, and additivity (or linearity). We show that these axioms can be justifiably used to characterize the proposed Shapley value-based voting rule.

\normalfont As mentioned already, our rule is simple: we model the voting rule by a TU game and then obtain the Shapley value for each alternative using the standard game-theoretic technique. The alternatives will be arranged in an order, we call this their collective ranking. From this order, we choose the top $k$ alternatives and form the required committee. Following the properties of the Shapley value, we show that in our proposed rule we are not only choosing the top $k$ alternatives from their collective ranking but choosing the top $k$ alternatives which have maximum compatibilities among themselves, as observed or believed by the voters. Recently, the Shapley value is used as a tool for ranking alternatives in (Moretti, $2015$); (Kondortev et al., $2017$), etc. The acceptance of the Shapley value as a suitable solution concept to different voting schemes can be seen in (Dehez et al., $2020$).\\
\normalfont Our evaluation process is based on a generalized version of approval by the voters. The classical approval voting introduced by (Brams et al., $1978$) is a special kind of evaluative voting with two levels of evaluation namely, approve or not approve. In practice, the voters are allowed to approve a list of alternatives. Among them, the one who receives the highest number of approvals becomes the winner. If it is a committee of size $k$, it is comprised of the $k$ alternatives with the highest number of approvals. There are several justifications for using approval ballots in multi-winner elections (i.e., to work with approval preferences). Compared to the ranking-based model, where voters provide complete rankings of alternatives (i.e., linear orders), providing approval preferences requires much less cognitive effort from the voters (Lackner et. al., $2022$). Thus, this kind of voting is often more practical and preferable due to its clear meaning. (Brams et. al., $2001$) and (Aragones et. al., $2011$) discuss the positive effects of using approval ballots on voters’ participation. Further, approval ballots are widely used in participatory budgeting (Goel et. al., $2015$). These are elections where the citizens decide through voting how to spend a municipal budget. The simplicity of the classical approval ballots also has downsides. An important underlying assumption is that the preferences of voters are separable, i.e., voters are not given the possibility to specify relations between alternatives. For example, in classical approval voting, it is not possible for a voter to indicate that she believes that a certain group of alternatives would work particularly well together in the elected committee or that she thinks that two alternatives should never be elected together. 

\normalfont In the classical approval voting, one can view each alternative as a singleton and any list of alternatives as a list of singletons. In this current paper, what we propose is an evaluation process where voters can approve a list of groups of alternatives. These groups will be of any size ranging from the singletons to the one with all the alternatives. In particular, when a voter is allowed to approve only a list of the singletons only, our evaluation becomes the classical approval voting. Thus, ours is a generalization of the classical approval voting.

\normalfont The rest of the paper proceeds as follows. In section \ref{sec:2}, we present the preliminary concepts. Section~\ref{sec:3}  describes a rule to compute a committee based on the Shapley value followed by its characterization using some standard axioms in section~\ref{sec:4}. Section \ref{sec:5} compares our committee rule with other committee selection rules. Finally, we conclude our paper in section \ref{sec:6}.
\section{Preliminaries}\label{sec:2}
\subsection{Compatibility-based committee selection framework} 
\normalfont Let $C = \{o_1, o_2,..., o_m\}$ be the set of alternatives with $|C| = m$ and $V = \{v_1, v_2,..., v_n\}$ the set of voters with $|V|= n$. The voters' evaluation scale is simply dichotomous. They have the ability to endorse multiple groups, ranging in size from $1$ to $m$. For simplicity, we denote $1$ as ``approved" and $0$ as ``not-approved". As mentioned in section~\ref{sec:1}, if a voter indicates approval of a particular group of size more than $1$, we assume that this voter believes that the alternatives belonging to that group are compatible.  Denote by $2^{|C|}$ the set of all groups of alternatives including the singletons. Let $E_{v} = (E_1,.., E_n)$ be the evaluation profile of $n$ voters. In particular, $E_{i}$ is the preference profile of voter $v_{i}$. Mathematically, $E_{i} \subseteq 2^{|C|}$. A rule $\Phi$ is a function that maps voters' evaluation matrix to an $m$ dimensional vector where $m$ is the set of alternatives. From this vector, we arrange them in an order based on some score assigned by the function $\Phi$. Thus, $\Phi$ is called a score function. Let $W$ denote the committee and $S_k(C)$ the set of all committees of size $k$. 
    
In what follows next, we present some intuitive properties that an ideal committee choice rule should satisfy. These properties, we compile from (Lackner et al., $2022$; Elkind et al., $2017$).
\subsection{Standard voting properties}
\begin{enumerate}[~~(a)]
\item \textbf{Unrestricted Domain(U):} Voters have the option to approve a list of groups of alternatives with size ranging from $1$ to $m$, i.e., the maximum size of a group with $m$ alternatives. 
\item \textbf{Anonymity(A):} All voters are treated equally. The permutations for names of voters do not change a collective ranking as well as the chosen committee.
\item \textbf{Neutrality(N):} All alternatives are treated equally. The permutations for names of alternatives do not change a collective ranking as well as the chosen committee.
\item \textbf{Monotonicity(M):} If alternative $o_1$ is in the chosen committee of size $k$, then $o_1$ would still be in the committee of size $k$ if some voters newly approve $o_1$  and that they did not approve $o_1$ earlier subject to the condition that other voters opinions are fixed in the evaluation profile.
\item \textbf{Inclusive property(I):} If an alternative is included in a chosen committee of size $k$, she must be included if the committee size is enlarged. (Elkind et al., $2017$) mention this property as shortlisting\footnote{Consider a situation where a position is filled at a university. Each faculty member ranks applicants in order to create a shortlist of those to be invited for an interview. One of the important requirements, in this case, is that if some alternative is shortlisted when $k$ applicants are selected, then this alternative should also be shortlisted if the list is extended to $k + 1$ applicants.}.
\end{enumerate}
At a later stage, we will show that the Shapley value-based committee rule (SV-CR) proposed in this paper, satisfies all the above-mentioned properties.
\subsection{Cooperative games with transferable utilities} Let $N=\{1, 2,...,|N|\}$ be the player set with $|N|$ players. In our terminology, $N$ denotes the set of alternatives. The subsets of $N$ are called coalitions and the set of all coalitions i.e., the power set of $N$ is denoted by $2^{|N|}$. To simplify notations, we write  $S\cup i$ for $S \cup\{i\}$ and $S\smallsetminus i$ for $S \smallsetminus \{i\}$ for each $S\subseteq N$ and $i\in N$. A cooperative game with transferable utilities or simply a TU game is the pair $(N, \delta)$ where the function $\delta: 2^{|N|}\rightarrow \mathbb{R}$ is such that $\delta(\emptyset)=0$. For each $S\subseteq N$, $\delta(S)$ denotes the worth of the coalition $S$. If $N$ is fixed, we denote a TU game by its characteristic function $\delta$ only. Denote $\delta_{0}$ the null game, defined as $\delta_{0}(S)=0$ for all $S\subseteq N$. The class of all TU games over the player set $N$ is denoted by $G(N)$ which forms a vector space of dimension $2^{|N|} -1$ under the standard addition and multiplications of set functions. For every coalition $S\subseteq N$ with $S\neq \emptyset$ , the game $e_{S}:2^{|N|}\rightarrow \mathbb{R}$ given by,
\[{e_{S}(T) =  \left\{ \begin{array}{ll}
& \mbox{$1 \hspace{.4cm}\textrm{if} \hspace{1mm} T = S$},\\
&\mbox{$0 \hspace{.4cm} \textrm{otherwise} $}\end{array} \right.}\]
is called the identity game. The identity game forms a standard basis for $G(N)$. Every element $\delta$ of $G(N)$ has a unique representation in of the identity games as  $\delta=\displaystyle\sum_{S\neq \emptyset}\delta(S)e_{S}$. Now let us define the marginal contribution of a player which is required in defining the concept of the Shapley value.
\begin{definition}
\normalfont {(Marginal contribution of a player, (Peters, $2008$)) For a TU game, $(N,\delta)$ an arbitrary player $i\in N$, and an arbitrary coalition $S$ that does not contain player $i$, the marginal contribution of a player $i\in N$ is defined as follows: 
$$m^{\sigma}(i) = \delta(S \cup i) - \delta(S).$$ Where $m^{\sigma}(i)$ is marginal contribution of player $i$. The expression $\delta(S\cup i)-\delta(S)$ describes the contribution of player $i$ when she joins coalition $S$.}  
\end{definition}

\normalfont A value is a function $\Phi : G(N) \rightarrow \mathbb{R}^{|N|}$ that assigns a payoff vector $\Phi(\delta)\in \mathbb{R}^{|N|}$ to each $\delta \in G(N)$ for a fixed player set $N$. The Shapley value (Shapley, $1953$) is given by the formula,
\begin{equation}\label{eq:shapley}
\Phi_{i}(\delta)=\sum_{S \subseteq N \smallsetminus i }\frac{|S|!(|N|-|S|-1)!}{|N|!}m^{\sigma}(i)
\end{equation}
For the characterization of the Shapley value given by Eq.(1), we require the following definitions.
\begin{remark}
    \normalfont The Shapley value determines the average of the marginal contribution of a player which measures how a player performs on an average in all the coalitions in the presence of the other players. In other words how a player performs in a group or committee. 
\end{remark}

\begin{definition}
\normalfont (Efficiency:) A value $\Phi: G(N) \rightarrow \mathbb{R}^{|N|}$ is said to be efficient if $\sum_{i \in N} \Phi_i(\delta) = \delta(N)$.
\end{definition}
\begin{definition}
\normalfont (Null Player:) A player $i\in N$ is a null player in $\delta$ if $\delta(S\cup i)=\delta(S)$ for all $S \subseteq N$.  The value $\Phi: G(N) \rightarrow \mathbb{R}^{|N|}$ satisfies the null player property if for each null player $i \in N$, $\Phi_i(\delta) = 0$.
\end{definition}
\begin{definition}
\normalfont (Symmetry:) Two players $i, j \in N$ are called symmetric with respect to the game $\delta$ if for all $S\subseteq N \smallsetminus \{i, j\}$, $\delta(S\cup i)=\delta(S\cup j)$. A value $\Phi:G(N) \rightarrow \mathbb{R}^{|N|}$ satisfies symmetry if for each pair of symmetric players, $\Phi_i(\delta) = \Phi_j(\delta)$.
\end{definition}
\begin{definition}
\normalfont (Linearity:) For $\delta, \gamma\in G(N)$ and real numbers $\alpha$ and $\beta$, the value $\Phi$ is linear if one has
$$\Phi(\alpha \delta + \beta \gamma) = \alpha \Phi(\delta) + \beta \Phi(\gamma).$$
In particular, for $\alpha = \beta =1$, the property $$\Phi( \delta + \gamma) = \Phi(\delta) + \Phi(\gamma)$$
is called additivity.
\end{definition}
\begin{theorem}\label{them:1}
\normalfont(Shapley, $1953$) The Shapley value given by Eq.(1) is the unique value that satisfies efficiency, null player property, symmetry, and additivity.
\end{theorem}

\normalfont There are many characterizations of the Shapley value. Here we state the characterization given by (Einy and Haimanko, $2011$). They characterize the Shapley value with additivity, dummy player, equal treatment of equal (equivalent to the symmetry), and gain-loss axioms. Before stating the theorem we define the dummy player and gain-loss axioms. 
\begin{definition}
\normalfont (Dummy Player:) A player $i\in N$ is a dummy player in $\delta$ if $\delta(S\cup i)=\delta(S) + \delta(i)$ for all $S \subseteq N$.  The value $\Phi: G(N) \rightarrow \mathbb{R}^{|N|}$ satisfies the dummy player property if for each dummy player $i \in N$, $\Phi_i(\delta) = \delta(i)$. {The dummy player is also known as a stand-alone player.}
\end{definition}

\begin{definition}
\normalfont (Gain-Loss:) For all $\delta, \gamma  \in G(N)$ and $i \in N$ such that $\delta(N)=\gamma(N)$ and $\Phi_{i}(\delta)>\Phi_{i}(\gamma)$, there is some $j \in N$ such that $\Phi_{j}(\delta)<\Phi_{j}(\gamma)$. 
\end{definition}
\noindent The Gain-loss axiom demands that whenever the size of a pie does not change, one player can only gain at the expense of the other (Casajus, $2014$). 

\begin{theorem}\label{them:2}
\normalfont(Einy and Haimanko, $2011$) The Shapley value given by Eq.(1) is the unique value that satisfies additivity, dummy, symmetry (``equal treatment for equals'' in their terminologies), and gain-loss.
\end{theorem}
\noindent We use these two characterizations of the Shapley value in the committee voting setup. 
\section{A TU game determined by the voters' generalized approval preferences}\label{sec:3}
\normalfont In this section, we formally define our model.
\normalfont Take $C = \{o_1,\cdots, o_m\}$ the set of $m$ alternatives and $V = \{v_1,\cdots,v_n\}$ the set of $n$ voters. For $E \subseteq 2^{|C|}$, define the generalized characteristic function $\chi_E : 2^{|C|} \mapsto \{0,1\}$ by
\begin{equation*}
    \chi_E(S) =\left\{\begin{array}{ll}
                   & 1 \;\;\textrm{if $S \in E$}\\
                   & 0 \;\;\textrm{otherwise}
           	       \end{array}\right.
\end{equation*}
Given the evaluation profile $E_v = (E_1,\cdots,E_n)$ of the $n$ voters, define the TU game $\delta: 2^{|C|}\rightarrow \mathbb{N} \cup \{0\}$, associated with the committee selection process based on $E_v$ by 
\begin{equation}\label{eq:TU}
	\delta(S) = \sum_{i=1}^n \chi_{E_i}(S)\;\;\textrm{for each $S \in 2^{|C|}$}.
\end{equation}
It follows from Eq.(\ref{eq:TU}) that for $S \in 2^{|C|}$, $\delta(S)$ gives the total number of approvals that group (coalition) $S$ receives in the voting process. Also, $\delta(\emptyset)=0$ and therefore, $\delta$ is a TU game. Call $\delta$ the associate TU game of $E_v$. In practice, the values given by $\delta$ over all possible groups (coalitions) of alternatives in $C$ can be calculated from Table~\ref{table:1} in the following, that represents the voters' complete evaluation profile by an $n \times 2^{m-1}$ matrix. The entries in the $i$-th  row will be either $1$ or $0$ according to the preference shown by voter $v_i$.
\begin{table}[h!]
\tabcolsep = 0.03cm
\begin{center}
\begin{tabular}{ |c|c|c|c|c|c|c|c|c| }
  \hline
  $\downarrow$ voters/alternatives $\&$ groups $\rightarrow$ & $\{o_1\}$ & $\{...\}$ & $\{o_m\}$ & $\{o_1, o_2\}$ &$\{...\}$$$ & $\{o_{m-1}, o_m\}$ & $\{..... \}$ & $\{o_1, o_2,... o_m\}$ \\
  \hline 
  $v_1$  & $\{0, 1\}$  & $\{0, 1\}$  & $\{0, 1\}$ & $\{0, 1\}$ & $\{0, 1\}$& $\{0, 1\}$ & $\{0, 1\}$ & $\{0, 1\}$ \\
  \hline
  $v_2$  & $\{0, 1\} $  & $\{0, 1\}$  & $\{0, 1\}$ & $\{0, 1\}$ & $\{0, 1\}$& $\{0, 1\}$ & $\{0, 1\}$ & $\{0, 1\}$ \\
  \hline
  ...    & ...     & ...     & ...    & ...    &... &...    &...     &...      \\
  \hline
  $v_n$  & $\{0, 1\}$  & $\{0, 1\}$  & $\{0, 1\}$ & $\{0, 1\}$ & $\{0, 1\}$ & $\{0, 1\}$ & $\{0, 1\}$ & $\{0, 1\}$ \\
   \hline
\end{tabular}
\caption{Voters' Complete Evaluation Profile Matrix}
\label{table:1}
\end{center}
\end{table}
\begin{definition}\rm
A committee selection rule is said to be a $k$-Shapley value-based committee rule ($k$-SV-CR) with respect to the associate TU game $\delta$ of a voters' complete evaluation profile if the committee is formed by including the first $k$-alternatives prescribed by the Shapley value of $\delta$.
\end{definition}
\begin{definition}\label{def:6}
\normalfont (Shapley Committee) A committee is said to be a Shapley Committee of size $k$ with respect to the associate TU game $\delta$ of a voters' complete evaluation profile $E_v$ if the committee members are the top $k$ alternatives from the collective rank given by the Shapley value-based committee rule associated with $\delta$.
\end{definition}

\noindent It follows from definition~\ref{def:6} that given a voters' evaluation profile $E_v$, if $W_1$ and $W_2$ are two committees of size $2$ say $W_1 = \{(x, y)\}$ and $W_2 = \{(a, b)\}$ then $W_1$ is a Shapley committee with respect to the associated TU game $\delta$ of $E_v$ if 
\begin{enumerate}[~~(a)]
\item $\Phi_x(\delta)\geq \Phi_y(\delta)$ or $\Phi_y(\delta)\geq \Phi_x(\delta)$ and 
\item $\min\{\Phi_x(\delta), \Phi_y(\delta)\} \geq \max\{\Phi_a(\delta), \Phi_b(\delta)\}$. 
\end{enumerate}

\begin{exm}\label{ex:4}\rm
\normalfont Now let us revisit Example \ref{ex:1} and Example \ref{ex:2}. Let's see if voters express their opinion correctly over groups including the singletons (individual alternatives), and whether this will reflect in our voting rule or not. For doing that we represent Example \ref{ex:1} in mathematical form. In Example \ref{ex:1}, there are three students $a$, $b$, and $c$ and we have to choose a group with two students from the voter's evaluations so that students are compatible with each other. For simplicity, we take four voters $v_1$, $v_2$, $v_3$, and $v_4$. However, one can take any number of voters. From Example \ref{ex:1}, it is clear that student $a$ is better than student $b$ and they do similar works. Let student $c$ have a different working skill, namely, the programming skill. Two voters namely, $v_1$ and  $v_4$ think that $a$ and $b$ both are equally good for the project and they approve both $a$ and $b$. However, out of the four voters, two voters namely, $v_2$ and $v_3$ only approve student $a$ for the project. And all four voters approve student $c$. These choices are based on the individual excellence of the alternatives known to the voters. Since voters can express their group excellency i.e., how students perform in a job together, we have the following evaluation profile by the voters. Out of four voters, two voters namely, $v_1$ and $v_2$ approve pair $(a, b)$ and the remaining two voters do not approve it. Similarly, only voter $v_1$ approves group $(a, c)$. But all the four voters approve group $(b, c)$ and group $(a, b, c)$. Voters' complete evaluations are illustrated in Table $2$.
\begin{table}[h!]
\begin{center}
\begin{tabular}{ |c|c|c|c|c|c|c|c|c| }
  \hline
  $\downarrow$ voters/alternatives $\&$ groups  $\rightarrow$ & $\{a\}$ & $\{b\}$ & $\{c\}$ & $\{a, b\}$ & $\{a, c\}$ & $\{b, c\}$ & $\{a, b, c\}$ \\
  \hline 
  $v_1$  & $1$  & $1$  & $1$ & $1$  & $1$ & $1$ & $1$ \\
  \hline
  $v_2$  & $1$  & $0$  & $1$ & $1$  & $0$ & $1$ & $1$ \\
  \hline
  $v_3$  & $1$  & $0$  & $1$ & $0$  & $0$ & $1$ & $1$ \\
  \hline
  $v_4$  & $1$  & $1$  & $1$ & $0$  & $0$ & $1$ & $1$ \\
   \hline
\end{tabular}
\caption{Evaluation Profile $E_v$ for Example \ref{ex:1}}
\label{table:2}
\end{center}
\end{table}
\end{exm}
\noindent With the help of the associate TU game $\delta$ of $E_v$, we obtain the following:   

$\delta\{a\} = 4$; $\delta\{b\} = 2$; $\delta\{c\} = 4$; $\delta\{a, b\} = 2$;  $\delta\{a, c\} = 1$; $\delta\{b, c\} = 4$; $\delta\{a, b, c\} = 4$.

\noindent This is the information on how voters approve alternatives in groups including the singletons (individual alternatives). Here $\delta\{a\} = 4$, describes that four voters approve alternative $a$ individually. Similarly, $\delta\{a, b\} = 2$, describes that $2$ voters prefer alternative $a$ and $b$ together in a committee. Once we have this information we can find the Shapley value for the alternatives and get the Shapley committee of size $k$ accordingly. For Example \ref{ex:1}, the Shapley value for each alternative is respectively $\Phi_{a}(\delta) = 0.83$; $\Phi_{b}(\delta) = 1.33$; $\Phi_{c}(\delta) = 1.83$. So, the collective ranking is $c \succ b \succ a$. Therefore, the Shapley committee of size $2$, for example, is $\{(c, b)\}$.

\normalfont In Example \ref{ex:2}, out of four voters three voters namely, $v_1$, $v_2$, and $v_3$ approve a bottle of water and a bottle of orange juice individually. Voter $v_4$ strictly approves only a bottle of water. But none approves sandwiches alone. However, their group acceptability among these three objects might be different from individual evaluations. One such situation we try to illustrate is the following. None of the voters prefers water and orange juice together for mountain hiking as they serve almost similar purpose and make the bag heavy together. However, they are all fine with the pairs: $(w, s)$, $(o, s)$, and $(w, o, s)$. This is illustrated in Table $3$. \\
\begin{table}[h!]
\begin{center}
\begin{tabular}{ |c|c|c|c|c|c|c|c|c| }
  \hline
  $\downarrow$ voters/alternatives $\&$ groups $\rightarrow$ & $\{w\}$ & $\{o\}$ & $\{s\}$ & $\{w, o\}$ & $\{w, s\}$ & $\{o, s\}$ & $\{w, o, s\}$ \\
  \hline 
  $v_1$  & $1$  & $1$  & $0$ & $0$  & $1$ & $1$ & $1$ \\
  \hline
  $v_2$  & $1$  & $1$  & $0$ & $0$  & $1$ & $1$ & $1$ \\
  \hline
  $v_3$  & $1$  & $1$  & $0$ & $0$  & $1$ & $1$ & $1$ \\
  \hline
  $v_4$  & $1$  & $0$  & $0$ & $0$  & $1$ & $1$ & $1$ \\
   \hline

\end{tabular}
\caption{Evaluation Profile $E_v$ for Example \ref{ex:2}}
\label{table:3}
\end{center}
\end{table}
\noindent From this information, the associated TU game of $E_v$ is given by the following. $\delta\{w\} = 4$; $\delta\{o\} = 3$; $\delta\{s\} = 0$; $\delta\{w, o\} = 0$;  $\delta\{w, s\} = 4$; $\delta\{o, s\} = 4$; $\delta\{w, o, s\} = 4$.\\
The Shapley value for $\delta$ is $\Phi_{w}(\delta) = 1.5$; $\Phi_{o}(\delta) = 1$; $\Phi_{s}(\delta) = 1.5$. Hence, the collective ranking is $w \sim s \succ o $. It follows that, the Shapley committee of size $2$ is $\{(w, s)\}$.

\normalfont Let us revisit Example \ref{ex:3} now. According to (Lackner et. al., $2022$) if we count by how many voters each alternative is approved, we see that $a$, $b$, and $c$ are approved more often ($66$ times). This can be seen as a good reason to choose the committee $\{(a, b, c)\}$; this committee contains the strongest alternatives. However, this committee essentially ignores the preferences of $34$ voters. Instead, one could choose the committee $\{(a, d, e)\}$, in which every voter finds one approved alternative. A more proportional committee would be $\{(a, b, d)\}$ or $\{(a, c, d)\}$ or $\{(b, c, d)\}$: here, $66$ voters approve $(a, b, c)$ which comprises roughly two-third of the population. 

\normalfont All these committees are reasonable. It is easy to justify on their behalf and to oppose them (Lackner et. al., $2022$). However, (Lackner et. al., $2022$) mainly discuss committees based on principles like individual excellence/ strongest alternative, diversity, and proportionality. But they ignore the compatibility issues among the alternatives that we are more concerned with. Now we present a hypothetical situation on Example \ref{ex:3} based on our approach. For simplicity, we consider only $10$ voters and five alternatives $a$, $b$, $c$, $d$, and $e$ and the chosen committee size is three i.e., $k = 3$. Since, in this case, we have five alternatives, therefore, we express voters' generalized approval evaluations in a slightly different way than the earlier one as defined in Table~\ref{table:1}. The way voters approve groups including the singletons (individual alternatives) are shown in Table~\ref{table:4}.  

\begin{table}[h!]
\begin{center}
\begin{tabular}{ |c|c|c| }
  \hline
  $\downarrow$ Voters   & $\rightarrow$ Alternatives $\&$ groups  \\
  \hline 
  $v_1$  & $\{a\}$; $\{a, c\}$  \\
  \hline
  $v_2$  & $\{a\}$; $\{b\}$; $\{a, c\}$   \\
  \hline
  $v_3$  & $\{a\} $\\
  \hline
  $v_4$  & $\{c\}$; $\{a, c\}$  \\
   \hline
  $v_5$  & $\{c\}$  \\
   \hline 
   $v_6$  & $\{e\}$  \\
   \hline
  $v_7$  & $\{e\}$  \\
   \hline  
    $v_8$  & $\{e\}$; $\{a, c\}$  \\
   \hline
  $v_9$  & $\{a\}$; $\{c\}$  \\
   \hline 
 $v_{10}$  & $\{a, c, e\}$  \\
   \hline
\end{tabular}
\caption{An Evaluation Profile for Example \ref{ex:3}}
\label{table:4}
\end{center}
\end{table}
\normalfont From this information, we get the associated TU game $\delta$ as follows. $\delta\{a\} = 4$; $\delta\{b\} = 1$; $\delta\{c\} = 3$; $\delta\{e\} = 3$; $\delta\{a, c\} = 4$; $\delta\{a, c, e\} = 1$. For all other coalitions/ groups $S \subseteq C$, $\delta\{S\} = 0$. The Shapley value for this game will be $\Phi_{a}(\delta) = 0.68$; $\Phi_{b}(\delta) = -0.48$; $\Phi_{c}(\delta) = 0.43$; $\Phi_{d}(\delta) = -0.73$ and $\Phi_{e}(\delta) = 0.1$. The collective rank will be $a \succ c \succ e \succ b \succ d$ and a committee of size three will be $\{(a, c, e)\}$.

\noindent Now we come to the second part of Example \ref{ex:3}. The two constraints mentioned in Example \ref{ex:3} that generate the particular voters' preferences over committees with three alternatives are $\{(a, c, d)\}$, $\{(a, c, e)\}$, $\{(a, d, e)\}$, $\{(b, c, d)\}$, $\{(b, c, e)\}$, $\{(b, d, e)\}$ $>$ $\{(c, d, e)\}$ $>$ $\{(a, b, c)\}$, 
$\{(a, b, d)\}$, $\{(a, b, e)\}$.

\normalfont As stated in (Uckelman, $2010$), given her preferences, how should this voter vote? If she votes for both Alice and Bob in hopes that only one will win, she risks electing one of her least-favored committees. If she votes for neither Alice nor Bob in hopes that other voters will prefer one over another, she risks electing her second choice committee $\{(c, d, e)\}$. If she votes for Alice but not Bob, or Bob but not Alice, she has no principled way to choose between these options, as she prefers Alice's committees and Bob-committees equally.

\normalfont This particular voter's problem arises from a wide variety of voting rules (Uckelman, 2010). It is hard to find a suitable voting rule, that this particular voter is satisfied with. However, in our rule, her problem can be solved at least partially. For example, she can approve
groups $(a, c, d)$ and $(b, c, d)$, without worrying about Alice and Bob together in the same committee. When she approves $(a, c, d)$ and $(b, c, d)$ and she is the only voter, she can get either a committee with Alice or a committee with Bob, but not both in the same committee. And this way she can vote without taking risks and simultaneously fulfill her wish. If every voter behaves exactly the same as this particular voter then the committee will be either Alice's committee with Charlie and Dave or Bob's committee with Charlie and Dave. Assume that there is only one voter and she approves groups $(a, c, d)$ and $(b, c, d)$. Then, the corresponding coalition form game will be $\delta(a, c, d) = 1$; $\delta(b, c, d) = 1$; and all $\delta(S) = 0$ for all $S \subseteq C$. The Shapley value for this game will be $\Phi_{a}(\delta) = -0.0167$; $\Phi_{b}(\delta) = -0.01678$; $\Phi_{c}(\delta) = 0.067$; $\Phi_{d}(\delta) = 0.067$ and $\Phi_{e}(\delta) = -0.1$. The collective rank will be $c \sim d \succ a \sim b \succ e$ and a committee of size three will be either $\{(c, d, a)\}$ or $\{(c, d, b)\}$. If there is another voter together with the previous voter and say this new voter approves group $(c, d, b)$ then the Shapley value-based committee with size three will be $\{(c, d, b)\}$ as then Shapley value for Bob is strictly greater than Alice. 

\section{Characterizations}\label{sec:4}
\normalfont In this section, we characterize the Shapley value-based committee selection rule using a set of intuitive axioms. These axioms are respectively efficiency or exhaustive property, additivity or consistency, symmetry or equal treatment for equals, and null player or neutral alternative. 
\begin{axiom}\normalfont ({\textbf{Exhaustive Property}})
\normalfont The score of the alternatives is an exhaustive allocation of the total evaluation. Mathematically, $\textrm{Score}(o_1)+ \textrm{Score}(o_2)+...+ \textrm{Score}(o_m)=\delta(o_1, o_2,.., o_m)$.
\end{axiom}
\noindent Here $\delta(o_1, o_2,.., o_m)$ i.e., the largest group (grand coalition in terms of the TU game) will be either positive or zero. If no voters approve the grand coalition then $\delta(o_1, o_2,.., o_m) = 0$. In that case, some alternative scores will be positive and some negative. Adding all of them makes zero i.e., exhaustive. However, this will not affect our rule as we just need to compare with alternatives. With a very mild restriction, our exhaustive property will be the same as the one-person-one vote property defined by (Dehez et. al., $2020$). We state the one-person-one-vote axiom formally.
\begin{axiom}\normalfont ({\textbf{One-person-one-vote, (Dehez et. al., $2020$)}})
\normalfont The scores add up to the number of voters or electors: $\textrm{Score}(o_1)+ \textrm{Score}(o_2)+...+ \textrm{Score}(o_m)= |V|$. 
\end{axiom}
\noindent Here $|V|$ is the number of voters. The restriction is that if every voter approves the grand coalition the above property will hold. See, Example \ref{ex:4}, where the sum of the alternatives score is equal to the number of voters i.e., $\Phi_{a}(\delta) + \Phi_{b}(\delta) + \Phi_{c}(\delta) = 0.83 + 1.33 + 1.83 = 4$ which is equal to the number of voters.

\normalfont Our first axiom i.e., the exhaustive property may be less appealing in case of the standard problems on voting. However, once we move on to voting on some resource allocation problems such as participatory budgeting, this property is very demanding. As resources must be utilized efficiently, they should be neither wasted nor left as surplus.

\normalfont Our third axiom is consistency. The requirement of consistency is easily adapted to the multi-winner case (Lackner et. al., $2021$). It says that if there are two groups of voters and for both of them the voting rule shows that committee $W_1$ is at least as good as committee $W_2$, then the rule must show that $W_1$ is at least as good as $W_2$ when the two groups join together in a single electorate. In the case of single-winner rules, this requirement is rather appealing. Rejecting it would be difficult to justify from the point of view of social philosophy as it would mean that we treat large and small societies differently (Lackner et. al., $2021$).
\begin{axiom}\normalfont ({\textbf{Consistency, (Lackner et. al., $2021$)}})
\normalfont A rule $\Phi$ is consistent if for each two profiles $E$ and $E^{'}$ over disjoint sets of voters, $V$ and $V^{'}$, and each of the two committees $W_1$, $W_2$ $\in S_k(C)$, $(i)$ if $W_1 \succ_{E}W_2$ and $W_1 \succeq_{E^{'}} W_2$, then it holds that $W_1 \succ_{E+E^{'}}W_2$, and $(ii)$ if $W_1 \succeq_{E} W_2$ and $W_1 \succeq_{E^{'}} W_2$, then it holds that $W_1 \succeq_{E + E^{'}} W_2$.
\end{axiom}
\noindent Our fourth axiom i.e., equal treatment for equals suggests that if the voters treat two alternatives equally, their score should be the same. This axiom carries the flavor of fairness rather than diversity or proportionality. Formally, we have the following.
\begin{axiom}\normalfont {\textbf{(Equal Treatment for Equals)}}
\normalfont If each voter approves two or more identical alternatives and groups, then their score will be the same. Alternatively, we can say those equally treated alternatives will make a tie in the collective rank.
\end{axiom}
\noindent The last axiom is neutral alternative. An alternative $o_{i}$ is called a neutral alternative if the number of voters approve of a group containing alternative $o_{i}$ together with other alternatives is equal to the number of voters who approve of a group with exactly the former group except alternative $o_{i}$ i.e., these two groups differ by only one alternative. Mathematically, for each $S \subseteq C \setminus o_i$, $S \cup o_i$ represents a group containing the alternative $o_{i}$ with other alternatives and $S$ represents a group with all alternatives in the former group except alternative $o_{i}$. If for all $S \subseteq C \setminus o_i$, $\delta(S \cup o_i) = \delta(S)$, then we call $o_{i}$ a neutral alternative. 

\begin{axiom}\normalfont {\textbf{(Neutral Alternative)}}
\normalfont The overall score of a neutral alternative is zero.
\end{axiom}
\begin{remark}
\normalfont Alternatives which are not neutral alternatives would either receive a positive score or a negative score. The alternative(s) with a positive score will come before the neutral alternative and alternative(s) with a negative score will come after the neutral alternative in the collective rank.     
\end{remark}

\begin{remark}
\normalfont It follows that a committee rule that satisfies all these properties ensures compatibility among the alternatives, gives fairness to the selection process, is indifferent to the equally treated alternatives, and finally exhausts all the scores of the voters. Moreover, the axiom’s neutral alternative and equal treatment for equals have their origins in the null player property and symmetry property of the Shapley value (Shapley, $1953$). The characterization of the $k$- SV-CR follows.
\end{remark}
\begin{theorem}
\normalfont The Shapley value-based committee selection rule is the unique evaluative voting rule that satisfies the exhaustive property, consistency, equal treatment for the equals and neutral alternative with respect to the TU game obtained by using the generalized approval-based evaluation by the voters.
\end{theorem}
\begin{proof}
It is not hard to show that $k$- SV-CR prescribed by Eq.(~\ref{eq:shapley}) satisfies the exhaustive property, consistency, equal treatment for equals, and neutral alternative axiom. On the other hand, the axioms: efficiency, additivity, symmetry, and null player property of the Shapley value are equivalent to the exhaustive property, consistency, equal treatment for equals, and neutral alternative of the $k$- SV-CR respectively. Therefore, by Theorem~\ref{them:1}, the $k$- SV-CR is the unique rule that satisfies the given axioms. 
\end{proof}
\newtheorem{prop}{Proposition}
\begin{prop}
\normalfont When every voter approves the grand coalition i.e., the group containing all the alternatives, the Shapley value-based committee selection rule is the unique rule that satisfies one-person-one-vote, consistency, equal treatment for equals, and neutral alternative property with respect to the TU game obtained by using the generalized approval-based evaluation by the voters.  
\end{prop}
\begin{proof}
The proof is straightforward. 
\end{proof}

\noindent Now, we proceed to our second characterization. The second characterization uses the gain-loss and dummy player property. First, we define gain-loss and dummy player axioms for committee selection setup.

\begin{axiom}\normalfont {\textbf{(Gain-Loss)}}
\normalfont  For all $\delta, \gamma \in G(C)$ and $o_{i} \in C$ such that $\delta(C) = \gamma(C)$ and $\Phi_{o_{i}}(\delta)>\Phi_{o_{i}}(\gamma)$, there is some $o_{j} \in C$ such that $\Phi_{o_{j}}(\delta)<\Phi_{o_{j}}(\gamma)$. 
\end{axiom}

\noindent The axiom of gain-loss implies that suppose there be two evaluation profiles, say $E$ and $E'$ by the same set of voters and accordingly we get two associate TU games, namely $\delta$ and $\gamma$ respectively. If one alternative is better off in $\delta$ over $\gamma$ then there is another alternative who is better off in $\gamma$ over $\delta$ whenever equal number of voters approve the whole group of alternatives in both the profiles. Thus, under this property, if an alternative $o_i$ is not included in a committee of size $k$ under the profile $E$ but included under $E'$, then there must be at least one alternative $o_j$ in $E$ who gets excluded from the committee under $E'$. Therefore, in a nutshell, we cannot make everyone better without hurting someone as long as the total space/seats/amount is fixed.\\
\normalfont We next propose the dummy or stand-alone alternative axiom. It connects the groups and singletons in the voters' approvals. The definition of the dummy player in the context of the committee selection process is as follows.\\
\noindent {An alternative $o_{i}$ is called a dummy or stand-alone alternative if, the number of voters that approve a group containing alternative $o_{i}$ is equal to the number of voters who approve a group not containing alternative $o_{i}$ plus the number of voters who individually approve that particular alternative $o_{i}$. Mathematically, for all $S \subseteq C \setminus o_i$, if $\delta(S \cup o_{i}) = \delta(S)+ \delta(o_{i})$, then $o_{i}$ will be the dummy/ stand-alone alternative. In the Shapley value-based voting rule, we assume that a dummy/ stand-alone alternative will get a score $\delta(o_{i})$ which is equal to the approvals of the singletons $\{o_{i}\}$. Suppose, no voters approve the singleton $o_{i}$.  Then we must have $\delta(S \cup o_{i}) = \delta(S)$ for all $S \subseteq C \setminus o_i$. Here, we call $o_i$ a neutral alternative. Thus, a dummy/ stand-alone alternative also refers to a neutral alternative if she is not approved by any voter individually.}
\begin{axiom}\normalfont {\textbf{(Dummy/ Stand-alone Alternative)}}
\normalfont {The overall score of a dummy/ stand-alone alternative is $\delta(o_{i})$ i.e., $\Phi_{o_{i}}(\delta)= \delta(o_{i})$.}
\end{axiom}

\begin{theorem}
\normalfont The Shapley value-based committee selection rule is the unique evaluative voting rule that satisfies the gain-loss property, consistency, equal treatment for the equals and dummy alternatives.
\end{theorem}
\begin{proof}
It is not hard to show that $k$- SV-CR prescribed by Eq.(~\ref{eq:shapley}) satisfies the gain-loss property, consistency, equal treatment for equals, and dummy alternative axiom. On the other hand, the axioms: gain-loss, additivity, symmetry, and dummy property of the Shapley value are equivalent to the gain-loss property, consistency, equal treatment for equals, and dummy-alternative of the $k$- SV-CR respectively. Therefore, by Theorem~\ref{them:1}, the $k$- SV-CR is the unique rule that satisfies the given axioms. 
\end{proof}
\noindent Now we state and prove a simple proposition that shows a connection between the $k$- SV-CR and $k$- approval voting. 
\begin{prop}
\normalfont If voters approve only singletons (individual alternatives) then the collective ranking of our rule ($k$- SV-CR) and  $k$- approval voting is the same. Similarly, if voters approve only groups of size $k$ then the committee chosen by the Shapley value-based rule is the committee whose group of size $k$ receives the highest approval.
\end{prop}
\begin{proof}
Suppose, $Ap(o_1) = \alpha_1$, $Ap(o_2) = \alpha_2$, ..., $Ap(o_m) = \alpha_m$ are the approval score of alternatives $o_1$, $o_2$,..., $o_m$. WLOG, we assume that $\alpha_1 > \alpha_2 >... > \alpha_m$. Then, collective ranking based on approval voting will be $o_1 \succ o_2... \succ o_m$. Therefore, a committee of size $k$ will be $\{(o_1, o_2,...,o_k)\}$. Now we will show that if voters only approve individual groups i.e., groups with only one alternative then Shapley value-based committee will coincide with the approval-based committee. Since voters only approve individual groups therefore we have the following. 
$$\delta(o_1) = Ap(o_1) = \alpha_1$$
$$\delta(o_2) = Ap(o_2) = \alpha_2$$ 
$$...$$
$$\delta(o_m) = Ap(o_m) = \alpha_m$$
\normalfont We will prove the result for three alternatives. From this, it will be clear that this will also hold for any number of alternatives. Let's assume the above conditions for three alternatives. 
$$\delta(o_1) = Ap(o_1) = \alpha_1$$
$$\delta(o_2) = Ap(o_2) = \alpha_2$$ 
$$\delta(o_3) = Ap(o_3) = \alpha_3$$
We have that $\alpha_1 > \alpha_2 > \alpha_3$. Now we express Shapley's value for all three alternatives.  
$$ \Phi_{o_1}(\delta) = \frac{1}{3!}\{2\alpha_1-\alpha_2-\alpha_3\}$$
$$ \Phi_{o_2}(\delta) = \frac{1}{3!}\{2\alpha_2-\alpha_1-\alpha_3\}$$
$$ \Phi_{o_3}(\delta) = \frac{1}{3!}\{2\alpha_3-\alpha_1-\alpha_2\}$$
\normalfont Let's check the ranking of the Shapley value. If possible, let $\Phi_{o_1}(\delta) < \Phi_{o_2}(\delta)$ then it implies $\alpha_1 < \alpha_2$ which is a contradiction. Similarly, we assume that, $\Phi_{o_1}(\delta) < \Phi_{o_3}(\delta)$ and $\Phi_{o_2}(\delta) < \Phi_{o_3}(\delta)$ then these two also implies $\alpha_1 < \alpha_3$ and $\alpha_2 < \alpha_3$ which are contradiction as we assume that $\alpha_1 > \alpha_2 > \alpha_3$. Then, the only relationship among them will be $\Phi_{o_1}(\delta) > \Phi_{o_2}(\delta) > \Phi_{o_3}(\delta)$, implies that $o_1 \succ o_2 \succ o_3$ coincides with approval based ranking. 

\noindent Second part of the proposition: When voters only approved groups of size $k$ then the approval-based committee coincides with the Shapley value-based committee. For three alternatives it is straightforward. Similarly, this can be shown for any number of alternatives but it will be lengthy and require the same steps many times. So, we omit it. 
\end{proof}

\begin{prop}
\normalfont In an unrestricted domain the Shapley value-based committee selection rule satisfies anonymity, neutrality, monotonicity, and inclusive property.  
\end{prop}
\begin{proof}
Anonymity, neutrality, and inclusive property are satisfied by the definition of the Shapley value. Monotonicity property can also be proven easily. Suppose, some voters change their decision to the approval of an alternative from not approval. Then, his score (Shapley score) is either increased or remains the same but not decreased which is standard in the Shapley formula. Further, in committee setup, it implies that if an alternative is included in a committee of size $k$ she will be in the committee if some voters approve that alternative from not approval. 
\end{proof}

\section{Comparisons with other committee rules}\label{sec:5}
Several committee selection rules have been developed recently. This also includes approval-based committee rules or shortly the ABC voting rules, for more details, see (Kilgour, 2010; Kilgour et. al., 2012; Lackner et al., $2022$). In this section, we compare our rule with some of the ABC voting rules. First, we compare our rule with the $k$-approval voting which is an excellence-based committee selection rule. With the examples already mentioned in this paper, we show how our rule is different from the $k$-approval voting. In Example \ref{ex:1}, the alternatives' scores based on the $k$-approval voting is respectively $Ap(a) = 4$; $Ap(b) = 2$; $Ap(c) = 4$. Hence, the collective rank will be $a \sim c \succ b$. Thus, the committee of size two is $\{(a,c)\}$. However, if we consider each group including the singletons (individual alternatives) and use the Shapley rule to identify the scores of alternatives then the Shapley scores are, $\Phi_{a}(\delta) = 0.83$; $\Phi_{b}(\delta) = 1.33$; $\Phi_{c}(\delta) = 1.83$. Hence, the collective rank will be $c \succ b \succ a$. Thus, a committee of size two is $\{(c, b)\}$ which is a more desirable committee in terms of the competence among the alternatives as perceived by the voters. In Example \ref{ex:2}, $Ap(w) = 4$; $Ap(o) = 3$; $Ap(s) = 0$. Hence, the collective rank will be $w \succ o \succ s$. The committee of size two is then $\{(w, o)\}$. Observe that it is a committee with two items that use almost the same purpose i.e., a bottle of water and a bottle of orange juice. If voters have some scope to provide group acceptability then the outcome might be more useful. Here, the scores of the alternatives based on Shapley value are respectively $\Phi_{w}(\delta) = 1.5$; $\Phi_{o}(\delta) = 1$; $\Phi_{s}(\delta) = 1.5$. Hence, the collective rank will be $w \sim s \succ o$, and the committee of size two should be $\{(w, s)\}$. This time we get a bottle of water and a sandwich. A more compatible committee. 

\normalfont Next, we compare our rule with the rule(s) where only group compatibility or group excellence are considered to find a committee of size $k$. One can argue that if we have to choose a committee of size $k$, why do we need to provide all the information which are not necessary? For example, consider the groups smaller or bigger than size $k$. How these combinations of alternatives can help in forming a committee of size exactly $k$? This is indeed a valid point to argue. However, when we only consider group compatibility or group excellence to find a committee of some specific size we lose or miss several good attributes and scopes. Now we illustrate this fact with Example \ref{ex:1}.

If we only consider the group score of the committees of size two then $\{(a, b)\}$, $\{(a, c)\}$ and $\{(b, c)\}$ are respectively $2$, $1$ and $4$. And their collective ranking will be $\{(b,c)\} \succ \{(a,b)\} \succ \{(a,c)\}$. And the winning committee will be $\{(b, c)\}$. Here, when we consider both i.e., individual excellence as well as group compatibility then also it generates $\{(b,c)\}$ as the best committee of size two. However, this is not always the same. For example, let us check it for Example \ref{ex:2}, here the committees of size two are $\{(w,o)\}$, $\{(w,s)\}$, $\{(o,s)\}$ and their scores are respectively $0$, $4$ and $4$. Here, committee $\{(w, s)\}$ and $\{(o, s)\}$ are tied i.e., when we only use group compatibility or group excellence there is no difference between them. However, if we use all those information i.e., individual excellence as well as group compatibility then results are different. When we consider individual excellence and group compatibility and use the Shapley rule, it produces committee $\{(w,s)\}$ as the best committee. This can be seen in Table $3$ as an individual bottle of water receives $4$ approvals whereas a bottle of orange juice receives only $3$ approvals. This makes a difference and generates a more clear outcome.

\section{Conclusions}\label{sec:6}
\normalfont The current paper introduces a new committee selection rule by adopting the properties of TU games.  Models involving TU games consider the interplay among the participating players under binding agreements. In many situations, voters are also concerned about this interplay among the alternatives while evaluating their acceptance and credibility as members of a committee. We show that providing very limited scope like dichotomous preferences (yes or no) indication from the voters in terms of both individual excellence and group competency makes the outcome significantly different which is one of the newnesses in our paper. The use of Shapley value as a committee voting rule gives a full ranking of alternatives, therefore, if we need a committee of size more (less) than $k$ we can easily include (remove) additional members in the committee. Two particular cases stated in proposition $2$ make our rule more general and trustworthy. The first is when voters only approve a singleton (individual alternatives) committee, then that committee is the same as the $k$- approval committee. Secondly, if voters approve only groups with size $k$ our rule produces a committee that has maximum approval votes in terms of groups with size $k$. There are many questions open to investigate committees that are based on individual excellence and group compatibility. One very interesting application is that of participatory budgeting. We can talk about how to choose $k$ compatible projects for society. This we keep for our future research work. 

\end{document}